\documentclass[12pt]{article}

\begin{document}

\newcommand{\be}{\begin{equation}}
\newcommand{\ee}{\end{equation}}
\newcommand{\bea}{\begin{eqnarray}}
\newcommand{\eea}{\end{eqnarray}}
\newcommand{\ack}[1]{[{\bf Pfft!: #1}]}

\newcommand{\eref}[1]{eq.\ (\ref{eq:#1})}
\def\NPB{{\it Nucl. Phys. }{\bf B}}
\def\PL{{\it Phys. Lett. }}
\def\PRL{{\it Phys. Rev. Lett. }}
\def\PRD{{\it Phys. Rev. }{\bf D}}
\def\CQG{{\it Class. Quantum Grav. }}
\def\JMP{{\it J. Math. Phys. }}
\def\SJNP{{\it Sov. J. Nucl. Phys. }}
\def\SPJ{{\it Sov. Phys. J. }}
\def\JETPL{{\it JETP Lett. }}
\def\TMP{{\it Theor. Math. Phys. }}
\def\IJMPA{{\it Int. J. Mod. Phys. }{\bf A}}
\def\MPL{{\it Mod. Phys. Lett. }}
\def\CMP{{\it Commun. Math. Phys. }}
\def\AP{{\it Ann. Phys. }}
\def\PR{{\it Phys. Rep. }}

\hyphenation{Min-kow-ski}
\hyphenation{cosmo-logical}
\hyphenation{holo-graphy}
\hyphenation{super-symmetry}
\hyphenation{super-symmetric}

\rightline{VPI-IPNAS-08-13}
\centerline{\Large \bf }\vskip0.25cm
\centerline{\Large \bf }\vskip0.25cm
\centerline{\Large \bf Non-relativistic AdS/CFT and Aging/Gravity Duality}\vskip0.25cm
\vskip .5cm

\centerline{{\bf Djordje Minic\footnote{dminic@vt.edu} and 
Michel Pleimling
\footnote{Michel.Pleimling@vt.edu}
}}
\vskip .5cm
\centerline{\it Department of Physics, Virginia Polytechnic Institute and State University,}
\centerline{\it Blacksburg, VA 24061, U.S.A.}


\begin{abstract}
We point out that the recent discussion of non-relativistic AdS/CFT
correspondence
has a direct application in non-equilibrium statistical physics,
the fact which has not been emphasized in the recent literature on the
subject. In particular, we propose a duality between aging in systems far from equilibrium characterized 
by the dynamical exponent $z=2$ and gravity 
in a specific background. The key ingredient in our proposal is the recent geometric realization of the 
Schr\"{o}dinger
group. We also discuss the relevance of the proposed correspondence for the more general
aging phenomena in systems where the value of the dynamical exponent is different from 2.
\end{abstract}


\newpage

\section{Introduction}

The problem of understanding the behavior of systems far
from equilibrium is one of the most challenging questions in contemporary many-body
physics \cite{michelreview}. In this context it turns out that local
scale invariance \cite{henkel94,henkel02} can play a very important role in understanding dynamical scaling
aspects of systems far from equilibrium \cite{michelreview,henkel06}.
In particular in the discussions of the phenomenon of aging \cite{michelbook}, in other words the
observation that properties of non-equilibrium systems generically depend on the
time since the system was brought out of equilibrium, the role
of the non-relativistic 
conformal group, i.e. Schr\"{o}dinger group, has been found to
be crucial in systems
with a dynamical exponent $z=2$ \cite{michelreview,henkel94,henkel02,henkel03,henkel04}.
In this note we propose that aging phenomena with the dynamical exponent $z=2$
can be understood from a dual, gravitational
point of view.

In this context it is useful to recall that 
recently there has been a lot of activity regarding the possible relevance of
the AdS/CFT correspondence to other difficult condensed matter problems such
as high temperature superconductivity \cite{adscft}.
The general concept of duality, on which a lot of recent progress in quantum field theory
and string theory is based, has also been used recently in various condensed matter settings 
\cite{sdfqhe}.
Most recently Son \cite{son} and independently Balasubramanian and McGreevy \cite{bal} have proposed
a geometrical realization of the Schr\"{o}dinger group
in what has been called \cite{son} ``an AdS/cold atoms correspondence''

In this note we wish to apply this geometric realization of the Schr\"{o}dinger group to an
aging/gravity correspondence.
We consider this as a first step in a new approach to understanding the
non-equilibrium dynamics of more general systems with $z \neq 2$,
and especially of disordered systems \cite{michelreview,henkel06,henkel08}.
On the one side we aim to bring the topic
of aging in systems far from equilibrium to the attention of string
and field theorists interested in applications of non-relativistic
AdS/CFT duality. Thus, section 2 briefly summarizes the importance of Schr\"{o}dinger
(non-relativistic conformal) group in aging in a self-consistent manner.
On the other side, we aim to bring the tools of non-relativistic
AdS/CFT to the attention of the physicists working in non-equilibrium
statistical physics, and in particular, in aging phenomena.
Sections 3 and 4 are devoted to them.
By putting together the information from these two areas of physics
we propose a duality between aging phenomena and gravitational physics
in specific backgrounds.
The ultimate aim of this duality is to be able to characterize 
different universality classes by computing the critical
indexes of relevant correlation functions as explained in
section 2, by using the classical physics of certain fields propagating
in gravitationally non-trivial backgrounds as explained in sections
3 and 4. Ultimately, the full non-perturbative correlation functions should be captured by 
the string theory \cite{recent3} sigma model in these backgrounds, thus bringing the
methods of string theory into non-equilbirum statistical physics and vice versa.

\section{Aging phenomena and the Schr\"{o}dinger group}

Let us start by briefly reviewing what is known about the dynamical scaling and scale invariance in
dynamical systems with aging \cite{michelreview,cugliandolo}.
The general set-up for the study of aging behavior is as follows:
One considers a coarse-grained order parameter $\phi(t,\vec{r})$, conjugate to a generalized
field $h(t, \vec{r})$, which is usually assumed to be fully disordered
at $t=0$, i.e. $\langle \phi(0, \vec{r} )\rangle = 0$. For a magnetic system, the
order parameter is of course the magnetization whereas the conjugate field is an external magnetic field.
In the following we consider the case where the order parameter is
not conserved by the dynamics (model $A$ dynamics). Typically, one studies the 
scaling behavior of two-time correlation functions (here and in the following we assume that
spatial translation invariance holds)
\be
C(t, s) = \langle \phi(t, \vec{r}) \phi(s, \vec{r}) \rangle \sim s^{-b} f_C(t/s)
\ee
as well as of two-time response functions
\be \label{scale1}
R(t,s) = \frac{\delta \langle \phi(t, \vec{r} ) \rangle}{ \delta h(t, \vec{r})} \sim s^{-1-a} f_R(t/s)
\ee
where $a$ and $b$ are non-equilibrium exponents whereas $f_C$ and $f_R$ are scaling functions.
This scaling behavior is expected in the aging regime, defined by both $t$ and $s$ as well as
$t-s$ being much larger than the characteristic microscopic time scale.
Note also that this scaling assumes a single characteristic length scale $L$ which
scales with time $t$ as
\be
L(t) \sim t^{1/z}
\ee
where $z$ is the dynamical exponent.
For large values of the argument one expects
\be \label{scale2}
f_C(x) \sim x^{-\lambda_C/z}, \quad f_R(x) \sim x^{-\lambda_R/z}
\ee
with new non-equilibrium exponents $\lambda_C$ and $\lambda_R$.
The aging behavior just summarized is called simple or full aging and has been observed in many
exactly solvable models, in numerical simulations of more complex models as well as in actual experiments.

Given the success of conformal invariance in equilibrium critical phenomena,
it is natural to ask whether scaling functions and the values of non-equilibrium exponents might
be deduced from symmetry principles by invoking generalized dynamical
scaling with a space-time dependent scale factor $b = b(t, \vec{r})$.
This program has been reviewed in \cite{michelreview} and here we concentrate on the
specific case when $L(t) \sim t^{1/2}$, i.e.
$z=2$. This is a very important case as it encompasses systems undergoing phase-ordering
with non-conserved dynamics \cite{bray}. The generalization for $z \neq 2$ has been discussed in \cite{michelreview}
and we will briefly comment on this in the concluding remarks of this paper.

The theoretical description of aging systems with a dynamical exponent $z=2$ starts from a stochastic
Langevin equation which for a non-conserved order parameter reads:
\be \label{langevin}
2 M \partial_t \phi = \nabla^2 \phi - \frac{\delta V[\phi]}{\delta \phi} + \eta
\ee
where $V$ is a Ginzburg-Landau potential and $\eta$ is a Gaussian white noise that arises due to the
contact with a heat bath \cite{footnote1}.
The theory of local scale invariance then permits to show \cite{michelreview, picone}
that under rather general conditions all averages (i.e., correlation and response functions) of the noisy theory 
(\ref{langevin}) can be reduced {\it exactly} to averages of the corresponding deterministic, noiseless theory.

For the $z=2$ case the relevant symmetry structure is the Schr\"{o}dinger group \cite{schrgrp}.
It is well known that the free diffusion equation 
\be
2 M \partial_t \phi = \nabla^2 \phi
\ee
is invariant under the Schr\"{o}dinger group (the free diffusion equation being essentially
equivalent to the free Schr\"{o}dinger equation).
The Schr\"{o}dinger group is defined through the space-time transformations
\be
t \to t' = \frac{\alpha t + \beta}{\gamma t+\delta}, \quad \vec{r} \to \vec{r'} = 
\frac{{\bf R} \vec{r} + \vec{v} t + \vec{a}}{\gamma t+\delta}
\ee
where $\alpha, \beta, \gamma, \delta$ and $\vec{v}, \vec{a}$ are real parameters and 
$\alpha \delta- \beta \gamma = 1$, whereas
$\bf{R}$ denotes a rotation matrix in $d$ spatial dimensions.
The algebra of generators of the Schr\"{o}dinger group consists of temporal
translations $H$, spatial translations $P^I$, Galilean transformations $\mathcal{K}^I$,
rotations $\mathcal{M}^{IJ}$, dilatations $D$ and the special conformal transformation $C$.
The algebra of generators of the Schr\"{o}dinger group \cite{michelreview, son} reads as follows:
\begin{eqnarray}
\left[\mathcal{M}^{IJ}, \mathcal{M}^{KL}\right] & = & i(\delta^{IK} \mathcal{M}^{JL}+\delta^{JL} \mathcal{M}^{IK}-
\delta^{IL} \mathcal{M}^{JK}-\delta^{JK} \mathcal{M}^{IL}), \\
\left[\mathcal{M}^{IJ}, P(\mathcal{K})^{K}\right] & = & i(\delta^{IK} P(\mathcal{K})^{J}-\delta^{JK} P(\mathcal{K})^{I})
\end{eqnarray}
where $P(\mathcal{K})$ denotes either the momentum $P$ or the Galilean boost $\mathcal{K}$ generator, and
\be
[D, P^I] = -iP^I, [D, \mathcal{K}^I] = i\mathcal{K}^I, [P^I, \mathcal{K}^I] = -i\mathcal{M}^{IJ}
\ee
and finally
\be
[D, H] = -2iH,   [C,H] = -iD, [D,C]=2iC
\ee
where $D$ simply rescales $t$ and $\vec{r}$ as $t \to e^{2 \rho} t$ and $\vec{r} \to e^{\rho} \vec{r}$
and $C$ acts as $t \to t/(1+\rho t)$ and $\vec{r} \to \vec{r}/(1+\rho t) $.

In complete analogy with the conformal field theory bootstrap \cite{polyakov}
one of the immediate consequences of Schr\"{o}dinger invariance is
a restricted form of the two- and three-point
functions for the $\phi$ fields \cite{michelreview}.
The two-point function is essentially given by the heat kernel (i.e. Green's function) 
of the diffusion equation (up to a normalization constant)
\be
\langle \phi_1(t_1, \vec{r}_1) \phi_2(t_2, \vec{r}_2) \rangle= \delta_{x_1, x_2} \delta_{M_1+M_2,0} 
(t_{1,2})^{-x_1} \exp(-\frac{M_1}{2} \frac{\vec{r}_{1,2}^{\,2}}{t_{1,2}})
\ee
where $t_{1,2} \equiv t_1-t_2$ and $\vec{r}_{12} \equiv \vec{r}_1 -\vec{r}_2$, with the 
scaling dimensions $x_1$ and $x_2$ and the masses $M_1$ and $M_2$.
Similarly the three-point function is 
\be
\langle \phi_1(t_1, \vec{r}_1) \phi_2(t_2, \vec{r}_2) \phi_3(t_3, \vec{r}_3)\rangle =
\delta_{M_1+M_2+M_3,0} 
(t_{1,2})^{-x_{12,3}/2}(t_{2,3})^{-x_{23,1}/2}(t_{1,3})^{-x_{13,2}/2} K
\ee
where $t_{i,j} \equiv t_i-t_j$, $\vec{r}_{ij} \equiv \vec{r}_i -\vec{r}_j$,
$x_{ij,k} \equiv x_i +x_j - x_k$
and where $K$ is given by
\be
 K \equiv\exp\left(-\frac{M_1}{2} \frac{\vec{r}_{1,3}^{\, 2}}{t_{1,3}}\right) \exp\left(-\frac{M_2}{2} 
 \frac{\vec{r}_{2,3}^{\,2}}{t_{2,3}}\right)
 F\left(\frac{(\vec{r}_{1,3}t_{2,3}-\vec{r}_{2,3}t_{1,3})^2}{t_{1,2}t_{2,3}t_{1,3}}\right)~.
\ee
Here $F$ denotes an arbitrary differentiable function.

These relations have an immediate application to aging: the response function is constrained and the
exponents follow. The correlation function is likewise constrained.
In deriving expressions for response and correlation functions one has to note that in the aging
regime time-translation invariance is broken and that only a sub-group of the Schr\"{o}dinger group that
does not contain time-translations has to be taken into account \cite{michelreview, picone}. It follows
that a quasiprimary scaling operator that transforms covariantly under the aging group is characterized by
the triplet $(x, \xi, M)$ where $\xi$ is a new ``quantum number'' associated with the field $\phi$ (see
\cite{michelreview, picone,hep} for details). In the field-theoretical setting the 
autoresponse function $R(t,s)$ is written as $R(t,s) = \langle \phi(t) \phi'(s) \rangle$ where the order parameter
$\phi$ and the  associated response field  $\phi'$ are characterized by the exponents $(x, \xi)$ and
$(x',\xi')$. From this one obtains the following expression for the autoresponse function \cite{michelreview}:
\be
R(t,s) = s^{-(x+x')/2}\left(\frac{t}{s} \right)^{\xi } \left( \frac{t}{s}-1 \right)^{-x - 2\xi} \Theta(t-s) \delta_{x+2\xi, x'+2\xi'}
\ee
where $\Theta(x)$ is the Heaviside step function. Comparing with the expected scaling behavior 
(\ref{scale1}) and (\ref{scale2}) yields for
the critical exponents (for $z=2$)
\be
\lambda_R  = 2(x+\xi ), \quad 1 + a = (x + x')/2~.
\ee
Similarly, for the spatio-temporal response function one obtains the expression
\be
R(t,s;\vec{r}) = R(t,s) \exp\left(-\frac{M}{2} \frac{\vec{r}^{\, 2}}{t-s}\right) ~.
\ee
Finally, explicit expressions can also be derived for the autocorrelation function \cite{henkel04,henkel07}.
These expressions are rather cumbersome and will not be reproduced here. 

It is important to note that all these predictions have been tested to yield the exact results in a large variety
of exactly solvable models and to describe faithfully over many time decades numerical data obtained
for more complex systems \cite{michelreview}.

\section{The Schr\"{o}dinger group and its associated geometry}

In this section we discuss an embedding of the Schr\"{o}dinger group in the conformal group
and the natural geometric realization of the Schr\"{o}dinger group as recently discussed by
Son \cite{son}.
We here follow the presentation by Son \cite{son} and Balasubramanian and McGreevy \cite{bal}, whereas 
an alternative derivation is discussed in
\cite{michelreview,henkel}.

Son \cite{son} and Balasubramanian and McGreevy \cite{bal} start 
from a manifestly conformally invariant massless Klein-Gordon equation in
(d+1)+1 dimensional Minkowski space time
\be
-\partial_t^2 \phi + \partial_i^2 \phi =0
\ee
where the summation of the repeated indices is assumed.
Using the light-cone coordinates
\be
x^{\pm} \equiv \frac{1}{\sqrt{2}} (x^0 \pm x^{d+1})
\ee
(a similar definition also holds for other quantities used below)
one can rewrite the massless Klein-Gordon equation as
\be
(- 2 \partial_{-} \partial_{+} + \partial_i^2) \phi =0 ~.
\ee
By identifying $\partial_{-} \equiv iM$ the Klein-Gordon equation becomes
the Schr\"{o}dinger equation with $x^{+}$ playing the role of time
\be
i \frac{\partial \phi}{\partial x^+} = - \frac{1}{2M} \partial_i^2 \phi ~.
\ee

The algebraic embedding of the generators of the Schr\"{o}dinger group into
the conformal group follows the embedding of the Schr\"{o}dinger equation into
the massless Klein-Gordon equation \cite{son,bal,michelreview,henkel}.
The conformal algebra is given by 
\be
[m^{IJ}, m^{KL}] = i(\eta^{IK} m^{JL}+\eta^{JL} m^{IK}-\eta^{IL} m^{JK}-\eta^{JK} m^{IL}),
[m^{IJ}, p^{K}]=i(\eta^{IK} p^{J}-\eta^{JK} p^{I})
\ee
where $m^{IJ}$ denote the rotation and $p^I$ the spatial translation generators
as well as 
\be
[\tilde{d}, p^I] = -ip^I, [\tilde{d}, k^I] = ik^I, [p^I, k^J] = -2i(\eta^{IJ} \tilde{d} + m^{IJ})
\ee
with dilatations $\tilde{d}$ and boosts $k^I$ and
$I,J,K,L =0,...,d+1$.
One now identifies the light cone momentum $p^+ = -i \partial_{-}$ with the non-relativistic mass $M$
as above, 
\be
p^{+} \equiv \frac{1}{\sqrt{2}} (p^0 + p^{d+1}) = M ~.
\ee
The operators that commute with $p^{+}$ then close into the Schr\"{o}dinger group
\be
H=p^{-}, P^I=p^I, \mathcal{M}^{IJ}=m^{IJ}, \mathcal{K}^I=m^{I+}, D=\tilde{d}+m^{+-}, 2C=k^+ ~.
\ee

Based on this algebraic embedding of the Schr\"{o}dinger group into the conformal group and
using the known geometric realization of the conformal group in $d+1$ dimensions in terms of the 
isometries of the Anti-de-Sitter
AdS$_{(d+1)+1}$ space
\be
ds^2= \frac{1}{u^2}( du^2 + \eta_{IJ} dx^I dx^J )~,
\ee
Son \cite{son} and Balasubramanian and McGreevy \cite{bal} 
have recently proposed the following metric which is 
invariant under $D= \tilde{d}+m^{+-}$
but not under the separate actions of $\tilde{d}$ or $m^{+-}$ 
as the natural geometric realization of the Schr\"{o}dinger group \cite{footnote2}
\be \label{z1}
ds^2 = -2 \frac{(dx^{+})^2}{u^{2z}} + \frac{-2dx^+dx^- +dx^idx^i +du^2}{u^2}~.
\ee
Note that this metric is not restricted to the case $z=2$ (the value of the 
dynamical exponent encountered in pure systems undergoing phase-ordering), but encompasses also cases
that are Galilean invariant and for which $z\ne 2$.

\section{Aging/gravity duality}

In this section we put the two previous sections together to propose
an aging/gravity duality. 
By the aging/gravity duality we mean a precise mathematical correspondence 
between aging phenomena and gravitational physics
in specific backgrounds which capture the geometry of the non-relativistic conformal group and 
its subgroups, such as the aging group in its simplest version.
The ultimate aim of this correspondence, as already emphasized in the
introduction, is to be able to compute the critical
indexes of relevant correlation functions 
by using the classical physics of certain fields propagating
in the gravitationally non-trivial background from section
3. Ultimately, the full non-perturbative correlation functions as well as 
characterization of different universality classes should be 
captured by the string theory in this background.

The dictionary we propose states that the
generating functional of 1PI correlation functions of certain operators $O$ relevant for
the physics of aging phenomena (in the non-relativistic CFT (NRCFT)) is equal to the
exponent of the action for certain fields $\varphi$ propagating in a geometric background, evaluated 
for the boundary values of these fields ($\varphi_b$)
equal to the sources $J$ for the operators $O$
\be
Z_{NRCFT} (J) = e^{-S(\varphi)}, \quad \varphi_b = J~.
\ee

For example, the relevant action for a scalar field $\varphi$ in the background 
discussed in section 3 is \cite{son, bal}
\be
S = \frac{1}{2} \int d^{d+3} x \sqrt{g}(\partial_I \varphi \partial_J \varphi g^{IJ} - m^2 \varphi^2)
\ee
where $d^{d+3} x \equiv d^d \vec{r} d t du dx^-$ ($x^+$ being the $t$ coordinate).
The equation of motion for $\varphi$ is
\be
\frac{1}{\sqrt{g}} \partial_I(\sqrt{g}g^{IJ} \partial_J \varphi) + m^2 \varphi =0 ~.
\ee
The solution ansatz is dictated by symmetries
\be
\varphi = f(u) e^{i\omega t + i \vec{k}\cdot \vec{r} + i M x^{-}}
\ee
and the radial differential equation for $f(u)$ is \cite{son, bal}
\be \label{z2}
[-r^{d+3} \frac{\partial}{\partial u}(r^{-d-1} \frac{\partial}{\partial u}) + 
(2l\omega +{\vec{k}}^2) r^2 + l^2 r^{4-2z} +m^2] f(u)=0
\ee
where close to the boundary $f(u) \sim u^{\Delta_{\pm}}$ with \cite{son, bal}
\be
\Delta_{\pm} = 1 + \frac{d}{2} \pm \sqrt{(1+ d/2)^2 + m^2 + \delta_{z,2} l^2}~.
\ee
Again, we give here the expression valid for general values of $z$.
Note that for $z=2$ and in $d=3$, which is relevant for the problem of aging in systems
undergoing phase-ordering \cite{footnote3},
\be
f(u) \sim u^{5/2} K_{\nu} (ku)
\ee
where $K_{\nu}$ is the modified Bessel function and 
\be
\nu = \sqrt{(5/2)^2 + M^2 +m^2}, \quad k^2 \equiv 2 M \omega + \vec{k}^2~.
\ee
By using the usual AdS/CFT dictionary \cite{ads}
one evaluates the on-shell action \cite{son, bal} to obtain (after introducing a cut-off near the
boundary parametrized by $x_b \equiv \vec{r}, t, x^{-}$ at $u=\epsilon$)
\be
S_0= \frac{1}{2} \int d^{d+2} x_b \varphi(x_b) \partial_u \varphi(x_b) 
\ee
which in momentum space gives
\be
\frac{1}{2} \int d p \varphi(-p) C(k, \epsilon) \varphi(p)
\ee
where 
\be
C(k, \epsilon) = \sqrt{g} g^{uu} f(r) \partial_u f(u)|_{u \to \epsilon}
\ee
with $f(u) \sim K_{\nu} (ku)$ so that 
the two point function of our order parameter $\phi$ (whose source in the functional integral is
represented by $\varphi$) is essentially given by $C(k, \epsilon)$:
\be
\langle \phi_1(\omega_1 \vec{k}_1)\phi_2(\omega_2 \vec{k}_2) \rangle = \delta(\vec{k}_1 + \vec{k}_2)
\epsilon^{-5}[(k^2 \epsilon^2)/4]^{\nu}
\ee
or in the position space 
\be
\langle \phi_1(t, \vec{r}) \phi_2(0, 0) \rangle= \delta_{\Delta_1, \Delta_2}  
(t)^{-\Delta_1} \exp\left(-\frac{M}{2} \frac{\vec{r}^{\, 2}}{t} \right)
\ee
which is precisely what we have found in section 2 based on the requirements of
the Schr\"{o}dinger invariance.

Thus the geometric interpretation and an AdS/CFT like dictionary obviously capture the
main symmetry constraints and thus also the result for the three-point function follows.
Consequently, the predictions for the scaling behavior in aging with $z=2$ follow as
well.

\section{Future directions}

In this note we have discussed a dictionary between aging phenomena and gravity based on
the geometric realization of the Schr\"{o}dinger group.
This discussion was based on the recent discussion of non-relativistic AdS/CFT duality
\cite{son, bal, other}.
Note that this dictionary is natural
from the point of view of the proposed closed relation between quantum gravity and
non-equilibrium statistical physics \cite{timem}.
This aging/gravity dictionary in some sense is an example that extrapolates the
Wilsonian dictionary between quantum field theory and equilibrium
statistical physics, to quantum gravity (i.e. string theory) in certain
backgrounds and certain non-equilibrium statistical mechanics phenomena.

What is a possible benefit that this geometric approach might have in the future 
regarding the more detailed understanding of aging in systems far from equilibrium?
We note that the example of phase-ordering kinetics encountered in the standard Ising
model, which is characterized by $z=2$ \cite{bray}, is already very interesting.
Nevertheless, the geometric approach does offer a possibility for 
treating systems whose dynamical exponent $z$ is different from 2.
It is apparent from the equations (\ref{z1}) and (\ref{z2}) 
that the geometric background dual to the Schr\"{o}dinger group can be discussed
for general $z$. Of course, the physics of systems with $z=2$ and $z\neq 2$ shows some differences
\cite{michelreview,henkel06,henkel08}, yet
the geometric picture does offer a new point of view on trying to understand the correlation
functions in the $z\neq 2$ case. Thus this geometrical picture might prove very useful in order
to incorporate into a unifying theoretical framework the recently extensively discussed superuniversality
of space-time quantities in disordered ferromagnets \cite{Sic08,henkel08}.

Furthermore, other order parameters can be considered even in the $z=2$ case.
One could obviously turn on the vector and tensor modes in the same background and
examine their correlators. Also, the higher order correlation functions, which can
be studied numerically, are amenable to the same geometric treatment.
Perhaps even a classification of different dynamical behaviors is possible in this case.
We plan to explore these issues in the future.

What is perhaps most interesting is that the geometric realization of the
Schr\"{o}dinger group can be extended to the aging group \cite{michelreview}, by considering
flows away from the non-relativistic conformal fixed point.
This is a familiar strategy in the domain of the AdS/CFT correspondence.
Note that the aging group in the simplest version is a subgroup of the
Schr\"{o}dinger group in which one gets rid of time-translational invariance, a necessary
requirement in order to describe non-equilibrium systems out of stationarity.

Finally, one should remember that ultimately one should be dealing with 
a string theory description in the backgrounds relevant for the Schr\"{o}dinger or
the aging group.
In that case, as in approaches to understand other strongly correlated systems such as
gauge theories, one ultimately has to deal with the dynamics of non-trivial two-dimensional
sigma models in non-trivial backgrounds.
Nevertheless, this new geometric viewpoint does open a new door in the field of aging
phenomena in systems far from equilibrium with many exciting and unforeseen applications.

\vskip .5cm

{\bf \Large Acknowledgements}

\vskip .5cm

We would
like to thank Uwe C. T\"{a}uber for a critical reading of the manuscript. {\small DM}
is supported in part by the U.S. Department of Energy
under contract DE-FG05-92ER40677.


\vskip 1cm


\begin{thebibliography}{99}

\bibitem{michelreview}
For a short review, see M. Henkel and M. Pleimling,
{\it Local scale-invariance in disordered systems}, 
in {\sl Rugged Free Energy Landscapes: Common Computational Approaches in
Spin Glasses, Structural Glasses and Biological Macromolecules}, editor W. Janke,
Lecture Notes in Physics {\bf 736}, 107 (Springer, Berlin, 2008);
cond-mat/0703466.
For a more exhaustive review, look at the book {\it Non-equilibrium phase transitions}, part II, by 
M. Henkel and
M. Pleimling, in preparation, and references therein.

\bibitem{henkel94} M. Henkel, 
J. Stat. Phys. {\bf 75}, 1023 (1994).

\bibitem{henkel02} M. Henkel, 
Nucl. Phys. B {\bf 641}, 405 (2002).

\bibitem{henkel06} M. Henkel and M. Pleimling, 
Europhys. Lett. {\bf 76}, 561 (2006).

\bibitem{michelbook}  M. Henkel, M. Pleimling, and R. Sanctuary (Editors), {\it Ageing and the Glass Transition},
Lecture Notes in Physics {\bf 716} (Springer, Berlin, 2007).

\bibitem{henkel03} M. Henkel and M. Pleimling, 
Phys. Rev. E {\bf 68}, 065101(R) (2003).

\bibitem{henkel04} M. Henkel, A. Picone, and M. Pleimling, 
Europhys. Lett. {\bf 68}, 191 (2004).

\bibitem{adscft}
S.~A.~Hartnoll, P.~K.~Kovtun, M.~Muller and S.~Sachdev,
  Phys.\ Rev.\  B {\bf 76}, 144502 (2007)
  [arXiv:0706.3215 [cond-mat.str-el]];
   S.~A.~Hartnoll and C.~P.~Herzog,
    Phys.\ Rev.\  D {\bf 76}, 106012 (2007)
  [arXiv:0706.3228 [hep-th]];
  S.~A.~Hartnoll, C.~P.~Herzog and G.~T.~Horowitz,
  arXiv:0803.3295 [hep-th];
   D.~Minic and J.~J.~Heremans,
  arXiv:0804.2880 [hep-th];
  S.~S.~Gubser and S.~S.~Pufu,
  arXiv:0805.2960 [hep-th].

\bibitem{sdfqhe}
For example,
C.~P.~Herzog, P.~Kovtun, S.~Sachdev and D.~T.~Son,
  Phys.\ Rev.\  D {\bf 75}, 085020 (2007)
  [arXiv:hep-th/0701036];
J.~J.~Heremans and D.~Minic,
  arXiv:0802.4117 [cond-mat.mes-hall]; V.~Jejjala, D.~Minic, Y.~J.~Ng and C.~H.~Tze,
  arXiv:0806.0030 [hep-th].
  

\bibitem{son}
D.~T.~Son,
  arXiv:0804.3972 [hep-th].

\bibitem{bal}
  K.~Balasubramanian and J.~McGreevy,
  arXiv:0804.4053 [hep-th] and references therein.
  
\bibitem{henkel08}
M. Henkel and M. Pleimling, 
arXiv:0807.1485 [cond-mat].

\bibitem{recent3}
 A.~Adams, K.~Balasubramanian and J.~McGreevy,
  arXiv:0807.1111 [hep-th];
  J.~Maldacena, D.~Martelli and Y.~Tachikawa,
    arXiv:0807.1100 [hep-th];
   C.~P.~Herzog, M.~Rangamani and S.~F.~Ross,
  arXiv:0807.1099 [hep-th].
 


\bibitem{cugliandolo}
  L. F. Cugliandolo, 
in {\it Slow relaxation and
non equilibrium dynamics in condensed matter}, editors
J.-L. Barrat, J. Dalibard, J. Kurchan, and M. V. Feigel'man (Springer, 2003).

\bibitem{bray} 
A. J. Bray, 
Adv. Phy. {\bf 43}, 357 (1994).

\bibitem{footnote1}
In some applications, notably in
the context of phase-ordering kinetics, it is useful to consider the following extension \cite{picone}
\begin{displaymath}
2 M \partial_t \phi = \nabla^2 \phi - \frac{\delta V[\phi]}{\delta \phi} - v(t) \phi + \eta
\end{displaymath}
where $v(t)$ is a time-dependent potential. Obviously, this equation can be reduced to the equation
(\ref{langevin}) through a gauge transformation.

\bibitem{picone}
A. Picone and M. Henkel, 
Nucl. Phys. B {\bf 688}, 217 (2004).

\bibitem{schrgrp}
H. A. Kastrup,
Nucl. Phys. B {\bf 7}, 545 (1968); 
C. R. Hagen,
Phys. Rev. D {\bf 5}, 377 (1972);
U. Niederer,
Helv. Phys. Acta {\bf 47}, 167 (1974).

\bibitem{polyakov}
A.~A.~Belavin, A.~M.~Polyakov, and A.~B.~Zamolodchikov,
  Nucl.\ Phys.\  B {\bf 241}, 333 (1984).

\bibitem{hep} M. Henkel, T. Enss, and M. Pleimling,
J. Phys. A {\bf 39}, L589 (2006).

\bibitem{henkel07} M. Henkel and F. Baumann, 
J. Stat. Mech. P07015 (2007).
  
\bibitem{henkel} M. Henkel and J. Unterberger, 
Nucl. Phys. B {\bf 660}, 407 (2003);
M. Henkel and J. Unterberger, 
Nucl. Phys. B {\bf 746}, 155 (2006).

\bibitem{footnote2}
This metric
has been thoroughly studied in a different context in \cite{gibbons}.

\bibitem{gibbons}
C.~Duval, G.~W.~Gibbons and P.~Horvathy,
  Phys.\ Rev.\  D {\bf 43}, 3907 (1991)
  [arXiv:hep-th/0512188] and references therein.
  
\bibitem{footnote3}
The $d=2$ situation
has been also treated in numerical studies. In that case obviously the geometric dual gives
$f(u) \sim u^2 K_{\sqrt{4+M^2+m^2}} (u \sqrt{2M\omega + {\vec{k}}^2})$.

\bibitem{ads}
J.~M.~Maldacena,
  Adv.\ Theor.\ Math.\ Phys.\  {\bf 2}, 231 (1998)
  [Int.\ J.\ Theor.\ Phys.\  {\bf 38}, 1113 (1999)]
  [arXiv:hep-th/9711200];
  S.~S.~Gubser, I.~R.~Klebanov and A.~M.~Polyakov,
    Phys.\ Lett.\  B {\bf 428}, 105 (1998);
    E.~Witten,
      Adv.\ Theor.\ Math.\ Phys.\  {\bf 2}, 253 (1998)
  [arXiv:hep-th/9802150]. For a review see,
  [arXiv:hep-th/9802109];
  O.~Aharony, S.~S.~Gubser, J.~M.~Maldacena, H.~Ooguri and Y.~Oz,
    Phys.\ Rept.\  {\bf 323}, 183 (2000)
  [arXiv:hep-th/9905111].
  
\bibitem{other}
See for example,
 W.~D.~Goldberger,
  arXiv:0806.2867 [hep-th];J.~L.~B.~Barbon and C.~A.~Fuertes,
  arXiv:0806.3244 [hep-th] and references therein.

\bibitem{timem}
V.~Jejjala, M.~Kavic and D.~Minic,
  Int.\ J.\ Mod.\ Phys.\  A {\bf 22}, 3317 (2007)
  [arXiv:0706.2252 [hep-th]];
 V.~Jejjala, M.~Kavic, D.~Minic and C.~H.~Tze,
  arXiv:0804.3598 [hep-th].

\bibitem{Sic08}
A. Sicilia, J.J. Arenzon, A.J. Bray and L.F. Cugliandolo, Europhys. Lett. 82, 10001 (2008).

\end{thebibliography}
\end{document}